# Remarks on the Woods-Saxon Potential

M. Çapak and B. Gönül

University of Gaziantep, Department of Engineering Physics, 27310 Gaziantep-Turkey

**Abstract**

More recently, comprehensive application results of approximate analytical solutions of the Woods-Saxon potential in closed form for the 5-dimensional Bohr Hamiltonian have been appeared [14] and its comparison to the data for many different nuclei has clearly revealed the domains for the sucsess and failure in case of using such potential forms to analyse the data concerning with the nuclear structure of deformed nuclei within the frame of the collective model. Gaining confidence from this work, exact solvability of the Woods-Saxon type potentials in lower dimensions for the bound states having zero angular momentum is carefully reviewed to finalize an ongoing discussion in the related literature and clearly shown that such kind of potentials have no analytical solutions even for $\ell = 0$ case.

## 1. Introduction

The shell model has been quite successful in describing nuclear structure, and the single-particle potential is important as the basic element of the shell model. For practical calculations, the single-particle potential is often assumed to be in a simple functional form, e.g. the Woods–Saxon (W-S) form [1] when compared to other more appropriate but complicated potentials such as the one used in Hartree-Fock calculations [2].

From the standpoint of a refined W-S functional form, the work in Ref. [3] has investigated the mathematical structure of a physically plausible single-particle potential which seems more accurate and applicable to a wider nuclidic region than attained before considering the related exhaustive study in Ref. [4]. The authors of Ref. [3] have clarified that the potential parameters for the extended W-S potential having a dip in the surface region constructed in their work are not independent from each other, which enables one to reduce the adjustable parameters. This point is significant in dealing with especially theoretical nuclear physics [5]. In the other words as the applications of this kind of interaction potential in nuclear structure analysis appear frequently in the literature, one needs to end up a wrong discussion ongoing in the previous related reports. The works in Refs. [6-11] on the W-S potential have argued that the standard form of this potential admits an exact solution for the vanishing angular momentum while providing approximate solutions for higher angular momenta within the frame of an appropriate scheme. In contrast to these reports, however, the work in Ref. [3] clearly proved that neither the spherically symmetric usual W-S potential nor its generalized form in there have a closed analytical solution even in one dimension for $\ell = 0$, which is also in agreement with the results of [12,13].

Within this context, the potential of interest has just recently been applied [14] to nuclear structure calculations in higher dimensions and observed that the corresponding new analytical solutions of the Bohr Hamiltonian, within the frame of collective model for the $\gamma-$unstable and $\gamma \approx 0$ rotational cases, well reproduce in particular the ground-state and $\gamma_1$ bandheads and energy spacings within the bands of many deformed nuclei as in agreement with the points posed in Ref. [3].

Therefore, a quick but clear discussion and a further justification of this point along [3, 14] is necessary for the further applications of such interaction forms within the framework of algebraically solvable Hamiltonians used in different models for not only the structure analysis of atomic nuclei but also the other fields in physics.

## 2. Discussion on the Woods-Saxon potential

This section deals with the analytic solution of the W-S potential within the consideration of non-relativistic physics. To clarify first that the standard form of this potential is not analytically soluable for its $s-$wave consideration, unlike some incorrect discussions in the literature [6-11], we start here with its extended form.

As the usual form of the W-S potential does not reproduce single-particle levels with enough accuracy when applied to a wide nuclidic region, extended forms of this potential are required for more physical applications, such as the ones in Refs. [3, 4]. In particular, the simplest form [3] of these functions has been frequently discussed [15-20] in the literature. For a clear understanding of the following discussion, we need at this stage to give a brief review of the related parts in Refs. [3, 14], which leads to the clarification of the two significant points that will be argued later in this section.

Reminding that the closed form of the algebraic solutions [3] for the modified spherically symmetric W-S potential,

$$V(r) = -\frac{V_0}{\left[e^{2a(r-R)}+1\right]} - \frac{We^{2a(r-R)}}{\left[e^{2a(r-R)}+1\right]^2} \tag{1}$$

are

$$E_n = -\frac{a^2}{4}\left\{\left[\sqrt{1+W/a^2}-(2n+1)\right]^2 + \left[\frac{V_0/a^2}{\sqrt{1+W/a^2}-(2n+1)}\right]^2\right\} - \frac{V_0}{2} \tag{2}$$

and

$$\Psi_n(r) = N\,\xi^{b/2}\left[1-\xi\right]^{c/2} P_n^{(b,c)}(1-2\xi) \tag{3}$$

where $\xi = \left[\dfrac{1}{e^{2a(r-R)}+1}\right]$ and $N$ is the normalization constant determined through $\int_0^\infty \left[\Psi_n(r)\right]^2 dr = 1$. In the above equations, $2a = 1/a_0$ with $a_0$ being the diffuseness parameter that determines the thickness of a surface layer in which the potential falls off from outside to its maximum strength inside a nucleus and $R$ is the mean radius of the nucleus of interest where the average interaction takes place. The additional term with $W$ in (1), which naturally arised in the calculations performed by [3], is responsible for modifying the standard well known W-S potential appeared as the first term in Eq. (1), providing the flexibility to construct the surface structure of the related nucleus.

The corresponding wavefunction in (3) is expressed in terms of the Jacobi polynomials due to the technique used in Ref. [3], in which $b, c > -1$, as the real parameters in the polynomial considered, which serves as a testing ground for the reliability of the present calculations and $n = 0, 1, 2, \ldots$ being as the radial quantum number. This significant point concerning with the certain mathematical and physically acceptable domain of the wavefunction parameters has missed from the attention of the related researchers [6-11] through their reports.

Along this line, the strengths of the potential in (1) can easily be expressed [3] in the form of

$$V_0 = a^2(b^2 - c^2) > 0 \quad , \quad W = a^2\left[(b+c)(b+c+4n+2) + 4n(n+1)\right] > 0 \tag{4}$$

which are the milestones in the forthcoming discussion given below in justifying that there is no explicit expression for the eigenvalues of the usual W-S potential in one- or three-dimensions. Clearly, from (4), one sees that $b + c = \left[\sqrt{1 + W/a^2} - (2n+1)\right] > 0$ and $b - c = \dfrac{V_0/a^2}{b+c} = \dfrac{V_0/a^2}{\sqrt{1+W/a^2} - (2n+1)}$, therefore the orthogonal polynomial parameters gain physical meanings such that

$$b = \frac{1}{2}\left[\sqrt{1+W/a^2} - (2n+1) + \frac{V_0/a^2}{\sqrt{1+W/a^2} - (2n+1)}\right]$$

$$c = \frac{1}{2}\left[\sqrt{1+W/a^2} - (2n+1) - \frac{V_0/a^2}{\sqrt{1+W/a^2} - (2n+1)}\right] \tag{5}$$

which impose some constraints on the potential parameters and certainly determines the number of physically reasonable bound states for a deep potential appearing near the surface as will be clear below in this section. In addition, these parameters clarify explicitly why the standart form of W-S potential in (1) cannot have a solution for $\ell = 0$ leading to $W = 0$ (through Eq. (11) belove), keeping of course the mathematical fact in the mind that $b, c > -1$ for the Jacobi polynomials. To our knowledege, such an analysis has never been appeared for the related problem in the earlier reports mentioned above.

Now, we are in a position to make an exciting connection between the attractive potential in (1) and the familiar effective W-S potential including the repulsive barrier term ($\hbar = 2m = 1$),

$$U_{eff}(r) = -\frac{U_0}{\left[e^{(r-R)/a_0} + 1\right]} + \frac{\ell(\ell+1)}{r^2} \tag{6}$$

which can not be exactly solved. Nevertheless, the recent works in [6-11, 13] have employed the widely used Pekeris approximation [21] in their calculations and replaced the angular momentum barrier term by

$$\frac{\ell(\ell+1)}{r^2} \approx \delta\left(C_0 + \frac{C_1}{e^{(r-R)/a_0} + 1} + \frac{C_2}{\left(e^{(r-R)/a_0} + 1\right)^2}\right) \tag{7}$$

taking into account the expansion terms up to the second-order, so that the $\ell-$ dependent effective potential in (6) preserves its original form, where

$$\delta = \frac{\ell(\ell+1)}{R^2} \quad , \quad C_0 = 1 - \frac{4a_0}{R} + \frac{12a_0^2}{R^2} \quad , \quad C_1 = \frac{8a_0}{R} - \frac{48a_0^2}{R^2} \quad , \quad C_2 = \frac{48a_0^2}{R^2} \tag{8}$$

Although this approach is valid only for low vibrational states, such a treatment does not cause any physical problem, which has been well discussed in [14], as this set of states contains the most important low-lying levels. Hence, the full potential in (6) is safely approximated to

$$U_{eff}(r) \approx \delta C_0 - \frac{(U_0 - \delta C_1)}{\left[e^{(r-R)/a_0} + 1\right]} + \frac{\delta C_2}{\left[e^{(r-R)/a_0} + 1\right]^2} \tag{9}$$

which, in its present form, yields algebraic closed solutions. Note that $\delta C_0$ constant in (9) will be later invoked into the corresponding energy value presented by Eq. (13).

To observe the close relation between the equation above and (1), we remind that $\frac{e^{2a(r-R)}}{\left[e^{2a(r-R)} + 1\right]^2} = \frac{1}{e^{2a(r-R)} + 1} - \frac{1}{\left[e^{2a(r-R)} + 1\right]^2}$, therefore Eq. (1) can easily be transformed to

$$V(r) = -\frac{(V_0 + W)}{\left[e^{2a(r-R)} + 1\right]} + \frac{W}{\left[e^{2a(r-R)} + 1\right]^2} \tag{10}$$

From the comparison of Eqs. (9) and (10) it is obvious that

$$W = \delta C_2 = \frac{12\ell(\ell+1)}{a^2 R^4} \quad , \quad V_0 = U_0 - \frac{4\ell(\ell+1)}{aR^3} \tag{11}$$

as $a_0 = 1/2a$ mentioned above. This simple analysis, unlike the previous works in Refs. [6-11], justifies clearly that both potential given by Eqs. (1) and (6) certainly do not admit an analytical solution in case $\ell = 0$ due to the additional strenght in the potentials $(W = 0)$ denoted by (11) that leads to unphysical polynomial parameters $(b, c)$ in (5) which should in fact satisfy $b, c > -1$ condition because of the mathematical definition of the Jacobi polynomials. Additionally, from (4), $b + c = \left[\sqrt{1 + W/a^2} - (2n+1)\right] > 0$ that also restricts the strengths of the physically acceptable potential parameters, especially for $\ell = 0$, as

$$0 \le n < \frac{\sqrt{1 + 12\frac{\ell(\ell+1)}{a^4 R^4}} - 1}{2} \tag{12}$$

in connection with the physically meaningful bound states exist in the system of interest. The reader is referred to Ref. [3, 14] for a more comprehensive discussion on this interesting constraint. A similar discussion on the same topic can also be found in [13].

To close this brief but necessary discussion, we go back to the solutions, Eqs. (2) and (3), of the modified W-S potential in (1) and bearing in mind the discussion above together in particular with the comparison results in (11), the replacement of the required terms reproduces explicitly the energy expression for the W-S potential in (9), being approximate form of (6), without solving the Schrödinger equation

$$\varepsilon_n = \frac{\ell(\ell+1)}{R^2}\left(1+\frac{12a_0^2}{R^2}\right) - \left[\frac{\sqrt{1+(192\ell(\ell+1)a_0^4/R^4)}-(2n+1)}{4a_0}\right]^2 -$$

$$\left[\frac{U_0 a_0 - 8\ell(\ell+1)a_0^2/R^3}{\sqrt{1+(192\ell(\ell+1)a_0^4/R^4)}-(2n+1)}\right]^2 - \frac{U_0}{2} \tag{13}$$

which is identical to the corresponding result obtained by [13]. The related wavefunction can be readily expressed, if necessary, in the same manner considering Eqs. (3), (5) and (11) in all together within the same frame. It is again obvious that $\ell = 0$ consideration yields unpysical results justifying the similar discussion above.

Furthermore, the generalized W-S potential in (10), or particularly its specific form namely the standard W-S potential in (9), reduces smoothly to a Morse-like potential

$$U_{eff}^{Morse}(r) \approx \delta C_0 - (U_0 - \delta C_1)e^{-(r-R)/a_0} + \delta C_2 e^{-2(r-R)/a_0} \tag{14}$$

in the $r > R$ domain when $a_0$ approaches to zero. Similarly, the same potential in (9) behaves like a finite quantum well, which is a familiar example in standard quantum text books, in the $R > r$ interaction region. These two considerations may be used for a further check of the present results.

Finally, we stress that the consideration of the same problem at large dimensions removes smoothly the mentioned drawback of $\ell = 0$ solution for the three dimensional Schrodinger equation with the standard Woods-Saxon potential as $\ell_N = \ell + \frac{N-3}{2}$ [13, 22] in this new case. The work in [14] carried out at $(N=)5$ dimensions verifies explicitly this interesting point.

## 3. Concluding Remarks

In this letter, exact solvability of spherically symmetric Wood-Saxon potentials for, in particular, $s-$wave consideration in three-dimension has been revised to clarify ongoing debates in the related literature. The present short discussion reveals clearly that neither the generalized W-S potential nor its standard form, discussed through Eq. (1), can have an analytical solution for the zero angular momentum at low dimensions due to the mathematical properties of the parameters concerning the corresponding wavefuntion expressed in terms of Jacobi polynomials [14]. Such an investigation concerning with the wavefunction parameters for the potential of interest has always escaped from the attentions of earlier works [6-11] in the related literature. However, considering the comprehensive discussion in [14], we remind that this problem can be easily removed by the increase of the dimensions. In this sense, the present results would be interesting not only for pure theoretical physicist but also for experimental physicists dealing with nuclear structure and reactions.

### Acknowledgments

The authors acknowledge the financial support of the Scientific and Technical Research Council of Turkey (TÜBİTAK) under the project number ARDEB/1002-113F218.